\documentclass[12pt,preprint]{aastex}

\slugcomment{Not to appear in Nonlearned J., 45.}

\shorttitle{The star cluster Whiting 1}
\shortauthors{Giovanni Carraro}

\begin{document}


\title{Whiting~1: a new Halo Young Globular Cluster}


\author{Giovanni Carraro}
\affil{Departamento de Astronom\'ia, Universidad de Chile, 
Casilla 36-D, Santiago, Chile}
\email{gcarraro@das.uchile.cl}

\altaffiltext{1}{Visiting Astronomer, Cerro Tololo Inter-American Observatory.
CTIO is operated by AURA, Inc.\ under contract to the National Science
Foundation.}
\altaffiltext{2}{Astronomy Department, Yale University, 
P.O. Box 208101, New Haven, CT 06520-8101 , USA}
\altaffiltext{3}{On leave from Dipartimento di Astronomia, Universit\`a di Padova,
Vicolo Osservatorio 2, I-35122, Padova, Italy.}


\begin{abstract}
We report on $BVI$ CCD photometry of a field centered  
in the region of the Galactic
star  cluster Whiting~1 down to $V=23.0$. 
This cluster has never been studied insofar, and we provide
for the first time estimates of its fundamental parameters,
namely, radial extent, age, distance and reddening.
Whiting~1 turns out to be a compact star cluster with
a diameter of about 1$^{\prime}$. We find that the cluster is
about 5 Gyr old and has a probable metal abundance around [Fe/H]=
-1.20. Its position at $b=-60^{o}.64$ and at a heliocentric
distance of about 45 kpc makes the cluster a rather strange object,
surely not a disk old open cluster, but perhaps the youngest
halo Globular Cluster insofar known.
\end{abstract}


\section{Introduction}
Whiting~1 (WHI B0200-03, l = 161.62$^{\circ}$,
b= -60.64$^{\circ}$; J2000 $\alpha$ = 2$^{\rm h}$02$^{\rm m}$57$^{\rm s}$,
$\delta$ = -3$^{\circ}$15'10'') was discovered by \citet{whi02}
during a search for Local Group Dwarf Galaxies in the Southern
Hemisphere. A 1200 secs V-band image (their Fig.~4) reveals that
this object is a star cluster very well resolved.
The authors comment that {\it it is a cluster of blue stars with
a distant galaxy cluster in the background}.
This is probably the reason for which Whiting~1 has later on
classified as a Galactic  open cluster (\citealt{dia02}).
However its faintness and the particular location in the Milky Way
in the anticenter direction at $b= -60^{o}.64$ cast some doubts on its open
cluster nature.\\
\noindent
In this paper we provide new photometric data with the aim to clarify
the cluster nature and to derive the first estimates of its
fundamental parameters.\\
\noindent
The layout of the paper is as follows. Sect.~2 illustrates  
the observation and reduction strategies. 
An analysis of  the geometrical
structure and star counts in the field of the cluster
are presented in Sect.~3, whereas a discussion of
the Color-Magnitude Diagrams (CMD) is performed in Sect.~4.
Sect.~5 deals with the determination of clusters reddening, 
distance and age and,
finally, Sect.~6 summarizes our findings. 
 
\section{Observations and Data Reduction} 
 
$\hspace{0.5cm}$
CCD $BVI$ observations were carried out with the eight CCD camera on-board
the  1. 0m telescope at Cerro Tololo Interamerican Observatory (CTIO,Chile), in the nights of 
December 15 to 16, 2004. 
With a pixel size of $0^{\prime\prime}.469$,  and a CCD size of 512 $\times$ 512
pixels,  
this samples a $4^\prime.1\times4^\prime.1$ field in the sky.\\
\noindent
The details of the observations are listed in Table~1 where the observed 
fields are 
reported together with the exposure times, the average seeing values and the 
range of air-masses during the observations. 
Figs.~1 shows a 900 secs I band image of the area of 
Whiting~1. In the figure East is up, and North on the right.\\

\noindent
The data have been reduced with the 
IRAF\footnote{IRAF is distributed by NOAO, which are operated by AURA under 
cooperative agreement with the NSF.} 
packages CCDRED, DAOPHOT, ALLSTAR and PHOTCAL using the point spread function (PSF)
method (\citealt{ste87}). 
The two nights turned out to be photometric and very stable, and therefore
we derived calibration equations for all the 130 standard stars
observed during the two nights in the \citet{lan92} 
  fields SA~95-41, PG~0231+051, Rubin~149, Rubin~152,
T~phe and    SA~98-670 (see Table~1 for details).

\noindent
The calibration equations turned out of be of the form:\\

\noindent
$ b = B + b_1 + b_2 * X + b_3~(B-V)$ \\
$ v = V + v_1 + v_2 * X + v_3~(B-V)$ \\
$ i = I + i_1 + i_2 * X + i_3~(V-I)$ ,\\

\noindent
where $BVI$ are standard magnitudes, $bvi$ are the instrumental ones and  $X$ is 
the airmass; all the coefficient values are reported in Table~2.
The standard 
stars in these fields provide a very good color coverage.
The final {\it r.m.s.} of the calibration are 0.049, 0.034 and 0.033 for the B, V and I filter, respectively.

\noindent
Photometric errors have been estimated following \citet{pat01}.
It turns out that stars brighter than  
$V \approx 22$ mag have  
internal (ALLSTAR output) photometric errors lower 
than 0.10~mag in magnitude and lower than 0.18~mag in color.
\noindent
The final photometric catalog for Whiting~1 (coordinates,
B, V and I magnitudes and errors)  
consists of 1757 stars and are made 
available in electronic form at the  
WEBDA\footnote{http://obswww.unige.ch/webda/navigation.html} site
maintained by J.-C. Mermilliod.\\

\section{Star counts and cluster size} 
\citet{dia02} report a preliminary estimate
of Whiting~1 diameter amounting to 1.2 arcmin. 
By inspecting Fig~1 we can recognize that Dias et al. estimate
is surely a reasonable one.\\
Since our photometry covers entirely the clusters area and part
of the surroundings, we performed star counts to obtain
an improved estimate of the clusters size.\\
We derived the surface stellar density by performing star counts
in concentric rings around the clusters nominal center
and then dividing by their
respective surfaces. Poisson errors have also been derived and normalized
to the corresponding surface. Poisson errors in the field star counts
turned out to be very small, and therefore we are not going to show them.
The final radial density profile for Whiting~1 is shown in Fig.~2
as a function of V magnitude.
The contribution of Galactic disk field has been estimated by considering
all the stars in the corresponding
magnitude bin, located outside 1.6 arcmin from the cluster center, 
and by normalizing counts over the adopted area.\\ 
The cluster seems to be populated  by stars of magnitude in the range
$18 \leq V \leq 24$, where it clearly emerges from the background.
In this magnitude range the radius is not larger than 0.5 arcmin.
\noindent
In conclusion,  we are going to adopt the  value of 0.5 arcmin as 
Whiting~1
radius throughout this paper. This estimate is in good agreement with
the value of 1.2 arcmin reported by \citet{dia02} for the cluster
diameter.

\section{The Colour-Magnitude Diagrams} 
In Fig. 3 we present the CMDs of Whiting~1 for all the detected stars.
In the left panel of the same figure we show the CMD in the 
$V$ vs $(B-V)$ plane, whereas in the right panel we show
the CMD in the $V$ vs $(V-I)$ plane.
These CMDs are not very easy to be interpreted, due to
the faintness of the stars.
However we can recognize a wide Main Sequence (MS) in both the CMDs,
extending from $V \approx 21$, where the Turn Off Point (TO) is located
down to $V \approx 23.5$.
The upper part of the CMD is more confused.  It seems that a Sub-giant
Branch (SGB) is actually present, and that the Red Giant Branch (RGB) starts
rising at $V \approx 20.8$, $(B-V) \approx 0.9$. The RGB however
is poorly populated. A group of stars at 
$18.0 \leq V \leq 18.3$, $(B-V) \approx 1.0$ may indicate the
presence of a clump.\\

\noindent
Better information can be derived from Fig.~4, where we show
the CMDs of Whiting~1 as a function of the distance from the cluster
nominal center. Here we consider only stars having errors in V-band
lower than 0.15 mag (see Sect.~2).
In the left panel all the stars are plotted, whereas in the middle panel
we only plot the stars which lie within the cluster radius (see Sect~3).
Finally, in the right panel we plot the stars outside the cluster
region, to show how the Galactic Field toward the cluster looks like.\\
Interestingly, the MS in the middle panel gets much thinner, and we can locate the TO at 
$V \approx 21.0$, $(B-V) \approx 0.5$. The RGB is scarcely populated
and the presumed clump disappears.
Most of the stars in the left panel located in the RGB region and
in the red edge of the MS come from the field
shown in the right panel.
\noindent
In conclusion, the overall shape of the CMD in the middle panel is 
reminiscent of an old cluster.

\section{Cluster fundamental parameters} 
In this section we provide estimates of Whiting~1
basic parameters. \\
We start deriving a first guess of the cluster reddening by using
FIRB maps by  \citet{sch98}.  We obtain E(B-V) = 0.026,
a very low value as expected for a cluster located at 
high Galactic latitude.\\
At this point, since we do not have any other information,
we have to rely on a detailed comparison between the CMD morphology
and theoretical isochrones. In the following analysis
we are going to adopt the Padova isochrones (\citealt{gir00}).\\
\noindent
In Fig.~5 and 6 we play with age and metallicity, trying to encounter
the best overall fit of the CMD.\\
In details, in Fig.~5 we keep the metallicity fixed to Z=0.001
and change the age from 3.8 to 5.6 Gyrs. In both the CMD the best fit
is provided by the age of 4.8 Gyrs (dashed-dotted line) which
implies a reddening E(B-V)= 0.04 and E(V-I)=0.05. The older isochrone
(5.6 Gyr) implies an untenable  E(B-V) = 0.00, or slightly negative,
while the youngest one (3.8 Gyr) implies a reddening E(B-V) = 0.12,
which is still acceptable, but the overall fit of the red part
of the CMD looks very poor.\\
On the other hand, in Fig.~6 we keep the age of 4.8 Gyr fixed and play
with metallicity. In this case the best fit is provided by Z=0.001.
The fit with a lower metal abundance is not so bad as well,
while the larger metallicity isochrone poorly fits the CMD
and implies a negative reddening.\\

\noindent
From this exercise we infer that the cluster metal abundance is around Z=0.001
and the age around 5 Gyrs. The best fit is therefore outlined in Fig.~7,
and from this we derive a distance modulus  $(m-M)_V$=18.4 and a reddening
E(B-V)=0.04. As a consequence, we obtain a heliocentric distance of 
about 45 kpc. The results we achieved are listed in detail in Table~3.

\section{Conclusions}
We have presented the first CCD $BVI$ photometric study of the 
star cluster Whiting~1. The CMDs we derive allow us to 
infer estimates of the cluster basic parameters, which
are summarized in Table~3.\\
\noindent
In detail, we find that:
 
\begin{description} 
\item $\bullet$ Whiting~1 is a compact star cluster with a radius of 0.5
arcmin;
\item $\bullet$ we propose that it is a 5 Gyr old  cluster with a low 
metal content Z = 0.001 ([Fe/H] $\approx$ -1.20);
\item $\bullet$ its position high onto the Galactic plane and its distance
from the Sun ($\approx$ 45 Kpc)  can hardly be reconciled with Whiting~1 being a disk old open cluster. 
\end{description}

\noindent
The combination of age, position and metallicity makes Whiting~1
a very puzzling object.\\
We can rule put the hypothesis  of an old open cluster, because
of the low metal content, and mostly beacuse of its positions.\\
The cluster bears some similarities with Palomar~1 (\citealt{ros98}),
an 8 Gyr old globular cluster, and other transitional
clusters like Terzan~7, Ruprech~106 and others mentioned in \citet{ros98}. \\
Following the kind of discussion 
by \citet{ros98}, we here tentatively propose that
Whiting~1 is the Milky Way youngest globular cluster insofar known,
although a clear explanation for its formation and evolution
remains very challenging.

\acknowledgments
The work of Giovanni Carraro is supported by {\it Fundaci\'on Andes}.
This study made use of Simbad and WEBDA databases.

\clearpage

\begin{figure}
\epsscale{.80}
\plotone{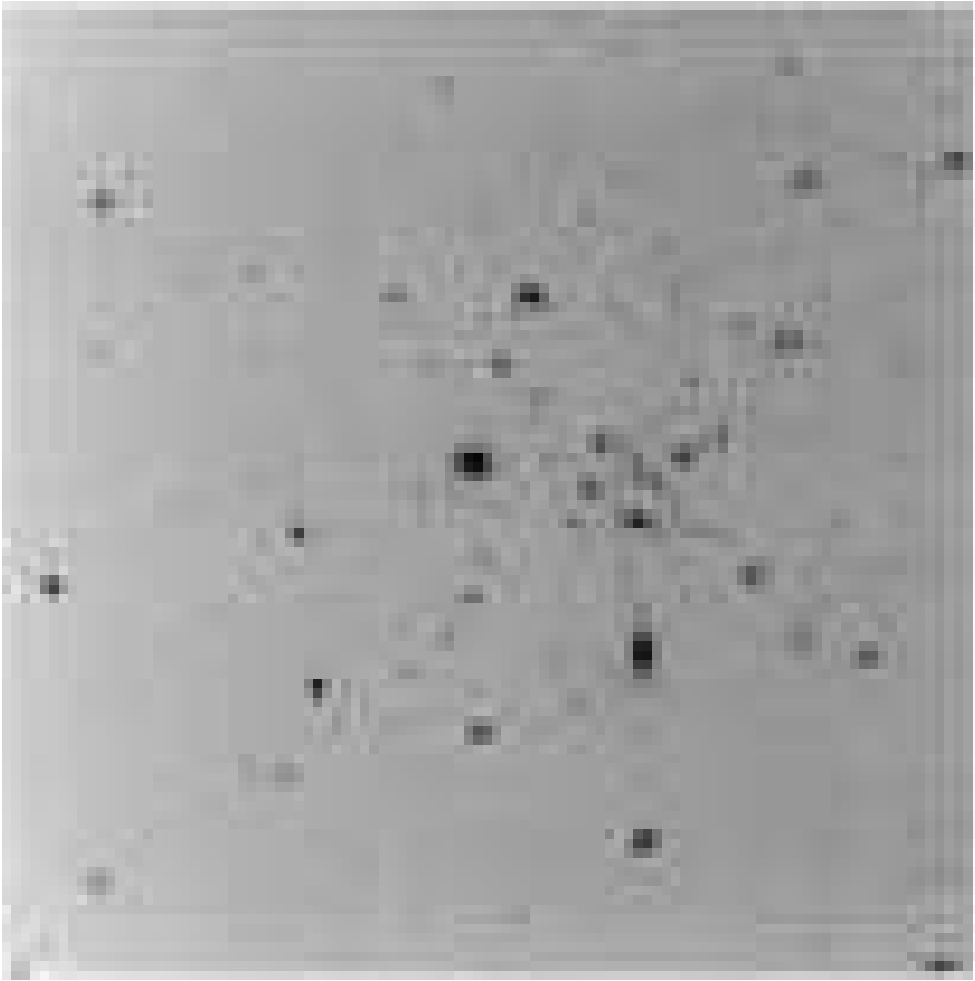}
\caption{900 secs I-band image of  the observed area in the region
of the open cluster Whiting~1. 
East is up, North on the right, and the covered area is $
4^{\prime}.1 \times 4^{\prime}.1$}
\end{figure}

\clearpage
\begin{figure}
\epsscale{.80}
\plotone{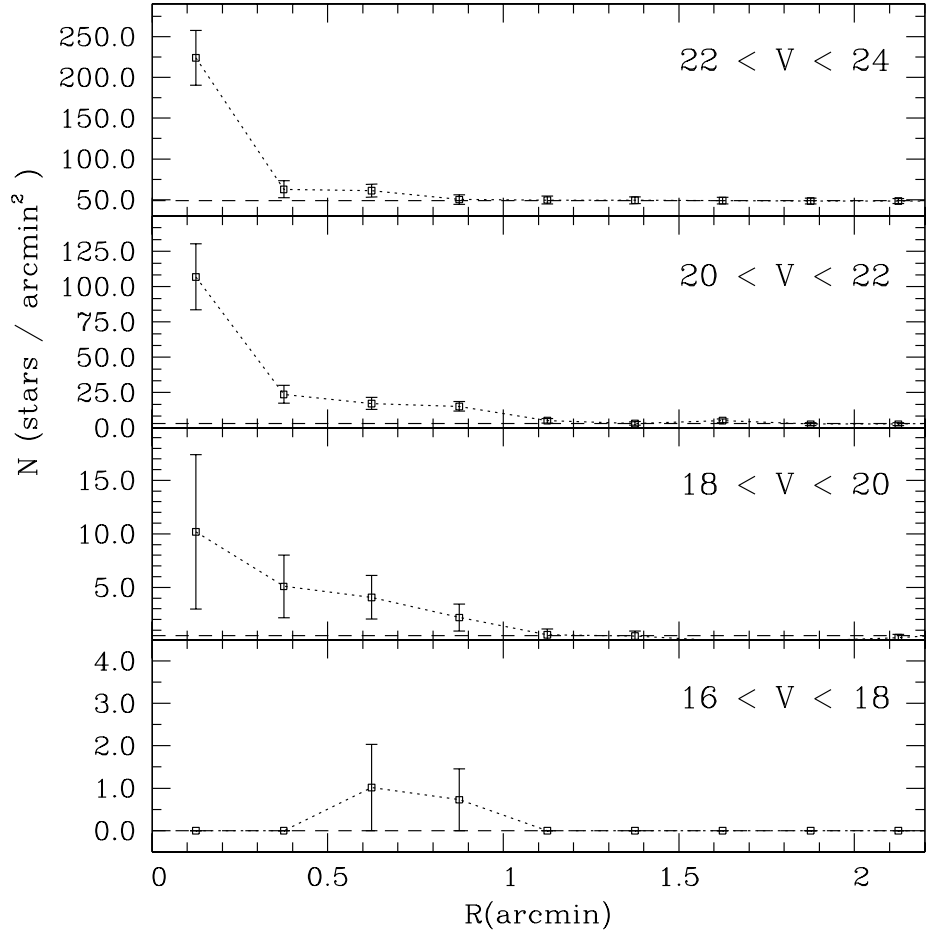}
\caption{Star counts in the area of 
Whiting~1 as a function of  radius and magnitude. The dashed lines represent
the level of the control field counts estimated in the surroundings
of the cluster in that magnitude range.}
\end{figure}

\clearpage
\begin{figure}
\epsscale{.80}
\plotone{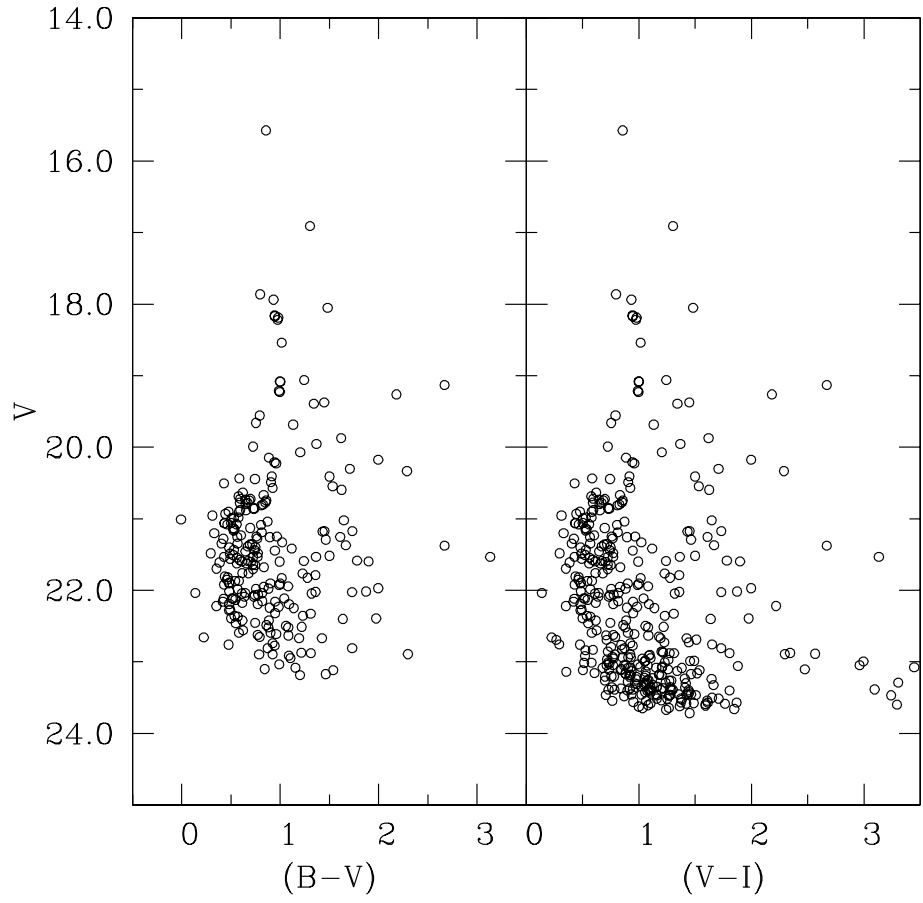}
\caption{$V$ vs $(B-V)$ (left panel) and $V$ vs $(V-I)$ (right panel) CMDs
of Whiting~1.}
\end{figure}

\clearpage
\begin{figure}
\epsscale{.80}
\plotone{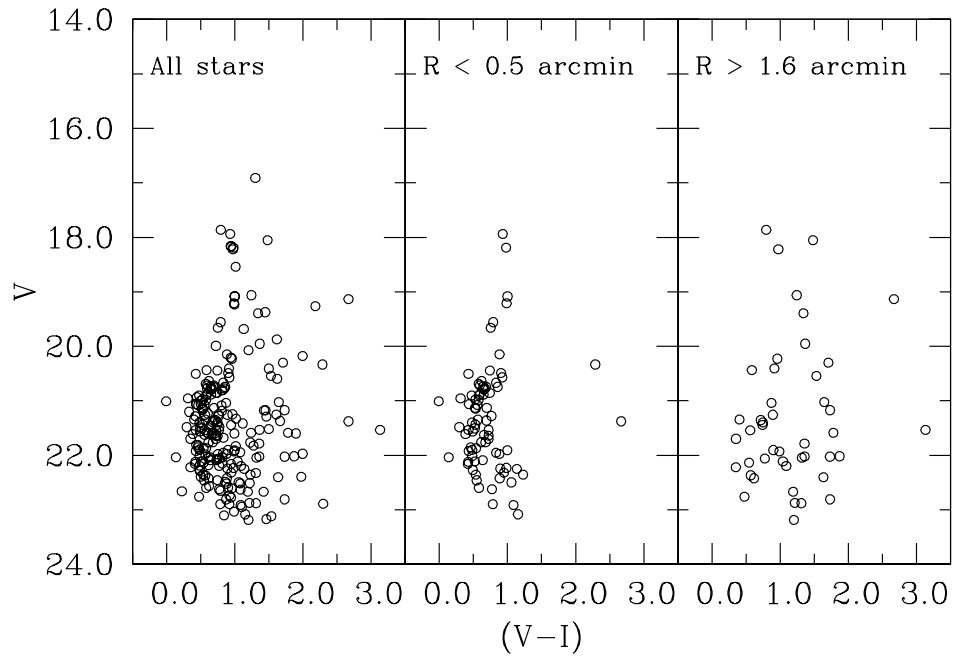}
\caption{$V$ vs $(V-I)$ CMDs
of Whiting~1 as a function of the distance from the cluster nominal 
center.}
\end{figure}

\clearpage
\begin{figure}
\epsscale{.80}
\plotone{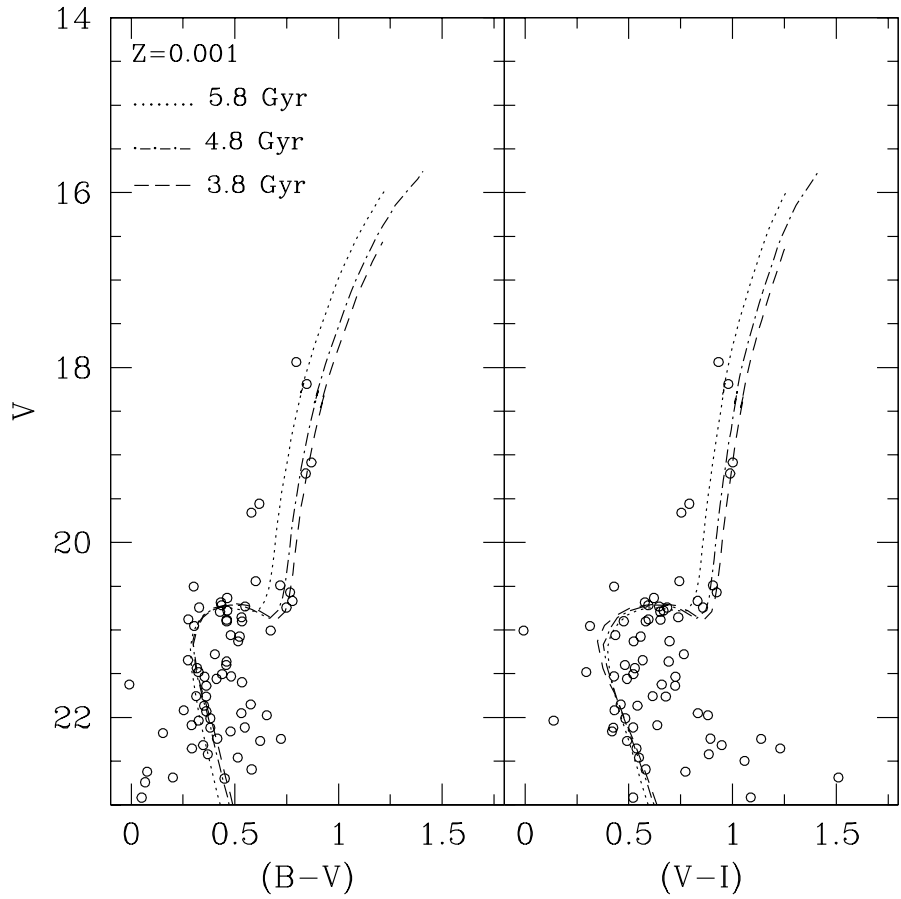}
\caption{Looking for the cluster age. Isochrone fitting
for the Z=0.01 metallicity. Dotted, dashed-dotted and dashed
isochrones are for the age of 5.8, 4.8 and 3.8 Gyrs, respectively.
See text for details}
\end{figure}

\clearpage
\begin{figure}
\epsscale{.80}
\plotone{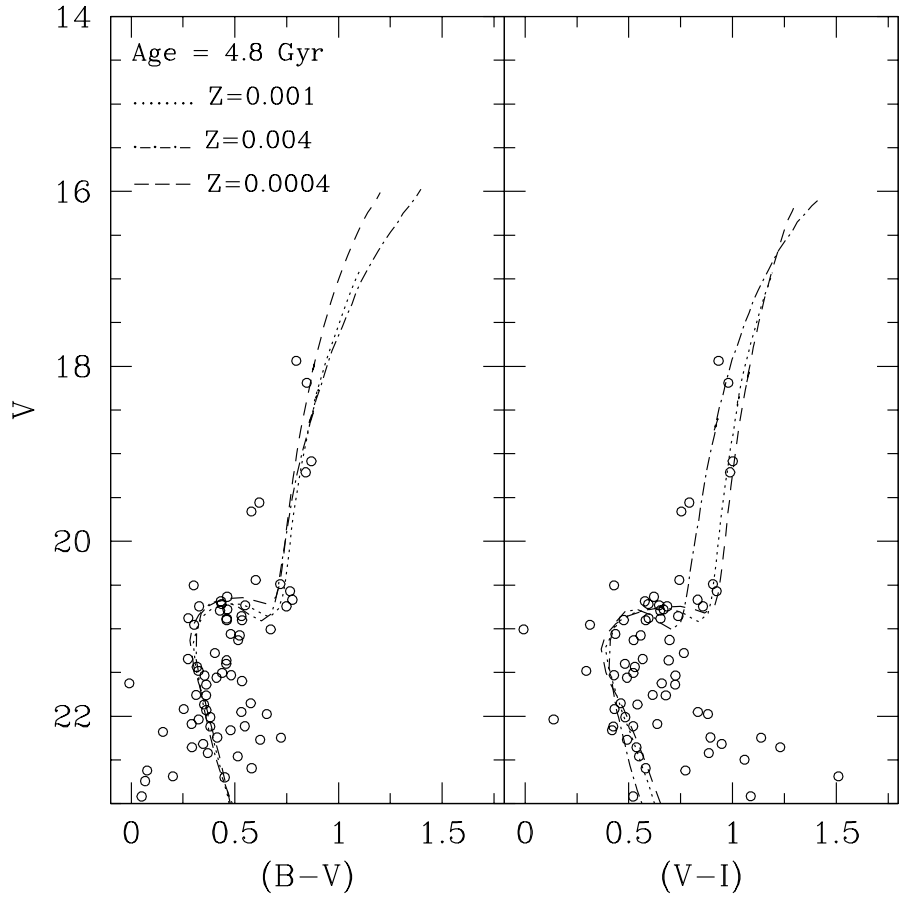}
\caption{Looking for the cluster metallcity. Isochrone fitting
for the the age of 4.8 Gyr. Dotted, dashed and dashed dotted
isochrones are for the metallicity Z = 0.001, 0.0004 and
0.004, respectively.
See text for details}
\end{figure}

\clearpage
\begin{figure}
\epsscale{.80}
\plotone{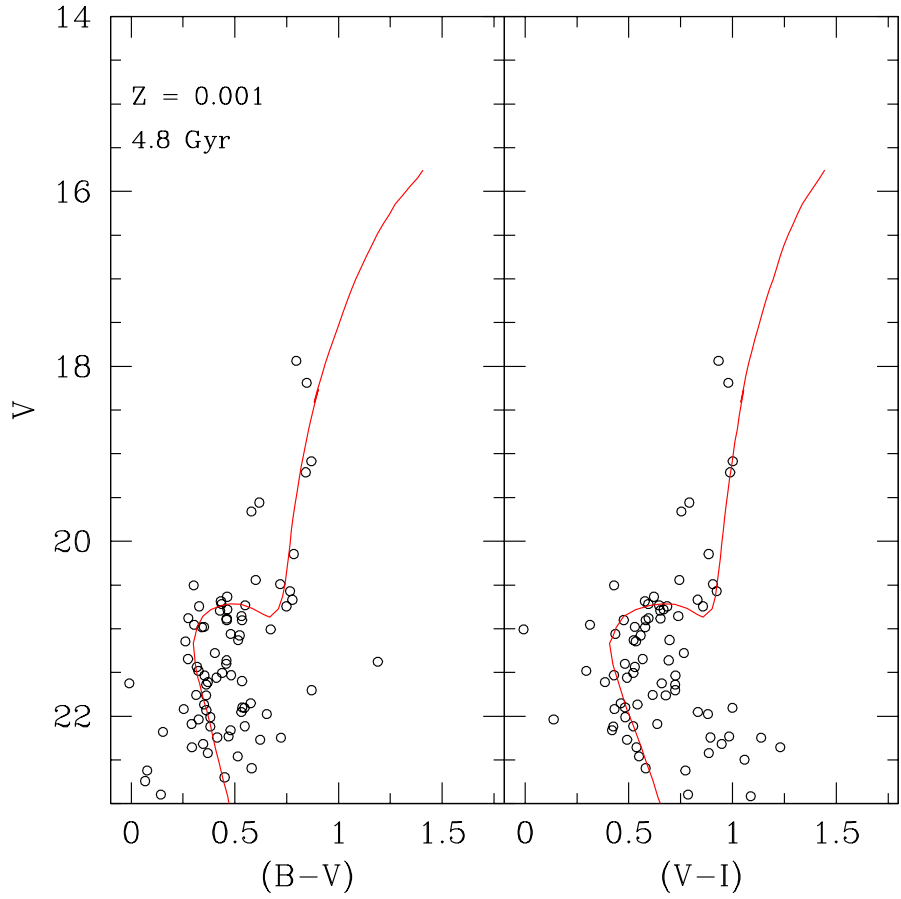}
\caption{Final isochrone solution for Whiting~1 CMD
obtained for the age of 4.8 Gyr, the reddening E(B-V)=0.04
and the distance modulus $(m-M)_V$=18.4. the isochrone is for 
a metallicity Z=0.001.}
\end{figure}

\clearpage
\begin{table} 
\fontsize{8} {10pt}\selectfont
\tabcolsep 0.10truecm 
\caption{Journal of observations of Whiting~1 
and standard star fields (December 15-16, 2004).} 
\begin{tabular}{cccccc} 
\hline 
\multicolumn{1}{c}{Field}         & 
\multicolumn{1}{c}{Filter}        & 
\multicolumn{1}{c}{Exposure time} & 
\multicolumn{1}{c}{Seeing}        &
\multicolumn{1}{c}{Airmass}       \\
 & & [sec.] & [$\prime\prime$] & \\ 
\hline 
Whiting 1     & B &     120,1200,1800   &   1.2 & 1.12-1.20 \\
              & V &      30,600,900    &   1.3 & 1.12-1.20 \\ 
              & I &      30,600,900    &   1.2 & 1.12-1.20 \\
\hline
SA 98-671     & B &   $3 \times$120   &   1.2 & 1.24-1.26 \\
              & V &   $3 \times$40    &   1.4 & 1.24-1.26 \\ 
              & I &   $3 \times$20    &   1.4 & 1.24-1.26 \\ 
\hline
PG 0231+051   & B &   $3 \times$120   &   1.2 & 1.20-2.04 \\
              & V &   $3 \times$40    &   1.5 & 1.20-2.04 \\ 
              & I &   $3 \times$20    &   1.5 & 1.20-2.04 \\ 
\hline
T Phe         & B &   $3 \times$120   &   1.2 & 1.04-1.34 \\
              & V &   $3 \times$ 40   &   1.3 & 1.04-1.34 \\ 
              & I &   $3 \times$ 20   &   1.3 & 1.04-1.34 \\ 
\hline
Rubin 152     & B &   $3 \times$120   &   1.3 & 1.33-1.80 \\
              & V &   $3 \times$40    &   1.2 & 1.33-1.80 \\ 
              & I &   $3 \times$20    &   1.2 & 1.33-1.80 \\ 
\hline
Rubin 149     & B &   $3 \times$120   &   1.1 & 1.21-1.96 \\
              & V &   $3 \times$40    &   1.2 & 1.21-1.96 \\ 
              & I &   $3 \times$20    &   1.2 & 1.21-1.96 \\ 
\hline
SA 95-41      & B &   $3 \times$120   &   1.2 & 1.05-1.48 \\
              & V &   $3 \times$40    &   1.2 & 1.05-1.48 \\ 
              & I &   $3 \times$20    &   1.1 & 1.05-1.48 \\ 
\hline
\hline
\end{tabular}
\end{table}

\clearpage
\begin{table} 
\tabcolsep 0.3truecm
\caption {Coefficients of the calibration equations}
\begin{tabular}{ccc}
\hline
$b_1 = 3.465 \pm 0.009$ & $b_2 =  0.25 \pm 0.02$ & $b_3 = -0.145 \pm 0.008$ \\
$v_1 = 3.244 \pm 0.005$ & $v_2 =  0.16 \pm 0.02$ & $v_3 =  0.021 \pm 0.005$ \\
$i_1 = 4.097 \pm 0.005$ & $i_2 =  0.08 \pm 0.02$ & $i_3 =  0.006 \pm 0.005$ \\
\hline
\end{tabular}
\end{table}

\clearpage
\begin{table*}
\caption{{}Fundamental parameters of Whiting~1.}
\fontsize{8} {10pt}\selectfont
\begin{tabular}{ccccccccc}
\hline
\multicolumn{1}{c} {$Radius(arcmin)$} &
\multicolumn{1}{c} {$E(B-V)$}  &
\multicolumn{1}{c} {$E(V-I)$}  &
\multicolumn{1}{c} {$(m-M)_0$} &
\multicolumn{1}{c} {$X(kpc)$} &
\multicolumn{1}{c} {$Y(kpc)$} &
\multicolumn{1}{c} {$Z(kpc)$} &
\multicolumn{1}{c} {$Age(Gyr)$} & 
\multicolumn{1}{c} {Metallicity}\\
\hline
 0.5 & 0.04 & 0.05 & 18.4 & 28.9 & 6.9 & -39.0 & 5.0& 0.001\\
\hline
\end{tabular}
\end{table*}

\end{document}